\documentstyle[12pt]{article}
\begin{document}
\vspace*{-1in}
\begin{flushright}
CERN-TH.7434/94 \\
TIFR/TH/94-33 \\
September 1994
\end{flushright}
\vskip 65pt
\begin{center}
{\Large \bf \boldmath A possible resolution of the
CDF $\psi^{\prime}$ anomaly}\\
\vspace{8mm}
{D.P.~Roy$^{*}$}\\
\vspace{5pt}
{\it Theory Group, Tata Institute of Fundamental Research,\\
Homi Bhabha Road, Bombay 400 005, India.}\\
\vspace{15pt}
{and}\\
\vspace{12pt}
{K. Sridhar$^{**}$}\\
\vspace{5pt}
{\it Theory Division, CERN, \\ CH-1211, Geneva 23, Switzerland.}\\

\pretolerance=10000

\vspace{70pt}
{\bf ABSTRACT}
\end{center}
We consider the contribution of radially excited
$2^3P_{1,2}$ states to $\psi^{\prime}$ production at
the Tevatron energy. Production of these states
$via$ the conventional gluon fusion mechanism and $via$ gluon and
charm quark fragmentation processes is considered. We find that it is
possible to account for the data on $\psi^{\prime}$ production
from the CDF experiment, by taking into account the decays of
these $2^3P_{1,2}$ states into $\psi^{\prime}$.
\vspace{40pt}
\noindent
\begin{flushleft}
CERN-TH.7434/94\\
September 1994\\
\vspace{11pt}
$^{*)}$ dproy@theory.tifr.res.in \\
$^{**)}$ sridhar@vxcern.cern.ch\\
\end{flushleft}

\vfill
\clearpage
\setcounter{page}{1}
\pagestyle{plain}
The CDF measurement \cite{cdf} of $J/\psi$ and $\psi^{\prime}$
production at large transverse momentum ($p_T$)in 1.8~TeV
$\bar p p$ collisions at the Tevatron is one of the
most challenging experimental results of recent times.
There is a huge discrepancy between the experimental
data and theoretical predictions of conventional models, much larger
than the uncertainties that are typical of such theoretical predictions.

\vskip10pt
The production of quarkonia at large $p_T$ is usually described by the
parton-fusion mechanism
in the framework of the colour-singlet model \cite{berjon,br,emc}.
In this model, the wave-function at the origin for a $S$-wave (or
its derivative for a $P$-wave) quarkonium is convoluted
with the cross-section for producing a heavy quark pair with
the proper spin, parity and charge-conjugation assignments.
There is an uncertainty in the determination of the wave-function
parameters, and taken in quadrature with that
due to choice of scale, an overall uncertainty
of a factor $\sim 3$ is expected in the predicted
normalisation of the quarkonium cross-section.
Within this normalisation uncertainty, the colour-
singlet model prediction agrees with the $J/\psi$ hadroproduction data
from fixed-target and ISR experiments over a reasonable range of $p_T$
\cite{br,schuler,gms}.

\vskip10pt
At higher energies where $b$-quark production becomes important, the
decay of $b$ quarks is a mechanism that contributes to $J/\psi$ and
$\psi^{\prime}$ production, in addition to the parton-fusion mechanism
of the colour-singlet model \cite{gms}. In the CDF experiment,
however, the secondary vertex information has been used to separate
out the $b$-quark decay contribution. The remainder are presumably
produced directly by the fusion mechanism: therefore, the discrepancy
between the large $J/\psi$ and $\psi'$ cross-sections measured by CDF
and the theoretical predictions shows that the fusion mechanism
grossly under-estimates the direct $J/\psi$ and $\psi^{\prime}$
production cross-sections.

\vskip10pt
The above discrepancy calls for a new mechanism of quarkonium
production at large $p_T$.  At high-energy colliders, where
gluons and charm quarks are copiously produced, the fragmentation
of these particles into quarkonium states has been suggested as
an important mechanism for large-$p_T$ quarkonium production
\cite{bryu}. Even though the fragmentation process is of higher order
in the strong coupling constant, $\alpha_s$, it can dominate over the
direct quarkonium production $via$ fusion at large $p_T$, where terms
of the order of $p_T^2/m_Q^2$ can easily compensate for the
suppression due to the extra power of $\alpha_s$. The validity of this
suggestion is borne out by the explicit computation of the fusion and
fragmentation contributions for the Tevatron energy, presented in
Refs.~\cite{ours, bdfm,cg}. In these papers, it was shown that the
fragmentation contribution is the dominant contribution at the
Tevatron energy; the largest contribution comes from gluons
fragmenting into $\chi$'s, which subsequently decay into $J/\psi$'s.
Together with the fusion contribution it can explain the large
cross-section measured by the CDF experiment to within a factor of
2-3, which is also the typical uncertainty in the predicted
normalisation.  In Ref.~\cite{ours}, the energy dependence of the
fragmentation and fusion contributions was studied as well. It was
shown that at ISR energies the fusion contribution dominates over the
$p_T$ range of interest, although the inclusion of the fragmentation
contribution helps to improve the quantitative agreement with the ISR
data.

\vskip10pt
In Refs.~\cite{ours,bdfm}, the production of $\psi^{\prime}$ by
fragmentation was also studied. Unlike $J/\psi$ production,
$\psi^{\prime}$ production does not involve the $P$-state
contribution. Consequently, the fragmentation contribution is
comparatively smaller. The total predicted $\psi^{\prime}$
cross-section is more than an order of magnitude smaller
than the cross-section measured by CDF.

\vskip10pt
In this letter, we explore the possibility that radially excited
$P$-state charmonia contribute $via$ their electromagnetic decays
to $\psi^{\prime}$ production. Some authors \cite{close,cho}
have recently suggested the possibility of a large
contribution of the $2^3P_{1,2}$
states to $\psi^{\prime}$ production at the Tevatron energy.
These $2P$ states have been predicted \cite{isgur1} to lie about
200-300 MeV above the open charm threshold, but their $D \bar D$
decay width is expected to be suppressed due to $L=2$ phase-space
as well as dynmical effects \cite{close,isgur2}.
Consequently, the branching ratios due to the decays $2^3P_{1,2}
\rightarrow 2^3S_1 + \gamma$ can be substantial, and, hence, the
$2P$ states could potentially contribute a significant fraction of
the measured $\psi^{\prime}$ cross-section \footnote{The contribution
of the ${}^1D_2$ state to the $\psi^{\prime}$ yield is expected to be
small. For a discussion, see Ref.~\cite{close}.} \cite{close,cho}.
We present the first quantitative estimate of this contribution.

\vskip10pt
The theoretical set-up which we need to compute the $p_T$ distributions
has been discussed in detail in Ref.~\cite{ours}. We will
compute the contributions from the fusion and the fragmentation
processes to the direct production of $\psi^{\prime}$
as well as to that of the $2P$ states.
The large-$p_T$ production cross-section for the fusion process is
given as
\begin{eqnarray}
\label{e1}
&&{d\sigma \over dp_T}(AB \rightarrow (2S,2P_{1,2}) X)
=  \nonumber \\
&& \sum \int dy dx_1 x_1G_{a/A}(x_1) x_2G_{b/B}(x_2)
{4p_T \over 2x_1 -\bar x_T e^y}
{d\hat \sigma \over d \hat t}(ab \rightarrow
(2S,2P_{1,2}) c) ,
\end{eqnarray}
where the sum, in the above equation, runs over all the partons
contributing to the subprocesses $ab \rightarrow (2S,2P_{1,2}) c$.
$G_{a/A}$ and $G_{b/B}$ are the distributions of the partons $a$ and $b$
in the hadrons $A$ and $B$ with momentum fractions $x_1$ and
$x_2$, respectively, and $x_2$ is given as
\begin{equation}
\label{e2}
x_2= {x_1 \bar x_T e^{-y} - 2 \tau \over 2x_1-\bar x_T e^y},
\end{equation}
where $\tau = M^2/s$, with $M$ the mass of the resonance, $s$
the centre-of-mass energy and $y$ the rapidity at which the resonance
is produced.
\begin{equation}
\label{e3}
\bar x_T= \sqrt{x_T^2 + 4\tau} \equiv {2M_T \over \sqrt{s}},
\hskip20pt x_T={2p_T \over \sqrt{s}}
\end{equation}
The expressions for the subprocess cross-sections, $d\hat\sigma/d\hat t$,
are explicitly given in Refs.~\cite{br} and \cite{gtw}.

\vskip10pt
The dominant fragmentation contribution comes from
the fragmentation of gluons and charm quarks.
It is computed by factorising the
cross-section for the process $AB \rightarrow (2S,2P_{1,2}) X$ into a
part containing the hard-scattering cross-section for producing a
gluon or a charm quark and a part which specifies the fragmentation of
the gluon (or the charm quark) into the required charmonium state,
i.e.
\begin{equation}
\label{e4}
d\sigma (AB \rightarrow (2S,2P_{1,2}) X)
 = \sum \int_0^1 dz \hskip4pt
d\sigma (AB \rightarrow c X) D_{c \rightarrow (2S,2P_{1,2})}(z,\mu ) ,
\end{equation}
where $c$ is the fragmenting parton (either a gluon or a charm quark)
and the sum in the above equation runs over the contributing partons.
$D(z,\mu)$ is the fragmentation function. The fragmentation function
which is computed perturbatively at an initial scale of the order of
the charm quark mass is evolved to the scale typical of the
fragmenting parton which is of the order of $p_T/z$, using the
Altarelli-Parisi equation.  For the production cross-sections of the
gluons and charm quarks we will use the lowest-order cross-sections,
$d\hat\sigma/d\hat t(ab \rightarrow cd)$, just as in the case of the
fusion contribution.  The full set of initial fragmentation functions
that we need to obtain the $2S$ and $2P$ contributions are~: $D_{g
\rightarrow 2S}$ \cite{bryu}, $D_{g \rightarrow 2P}$ \cite{bryu2},
$D_{c \rightarrow 2S}$
\cite{brcyu} and $D_{c \rightarrow 2P}$ \cite{chen,yuan}.
The final cross-section for the fragmentation process is given by a
formula similar to Eq.~\ref{e1} but with an extra integration
over $z$, or equivalently over $x_2$.

\vskip10pt
As mentioned earlier, a major uncertainty in these calculations is due
to the wave-functions. In the computation of the fusion contribution
these appear in the subprocess cross-sections, whereas for the
fragmentation contribution, the fragmentation functions at the initial
scale are proportional to the wave-function factors.  The wave
function factor $R_0^2$ for $\psi^{\prime}$ is reasonably
well-determined \cite{schuler}, but we have to make assumptions for
the parameters that describe the $2P$ states.  In Ref.~\cite{bryu2}
the fragmentation function for $g\rightarrow \chi$ is written in terms
of two parameters $H_1$ and $H_8^{\prime}$, where $H_1$ is related to
$R_1^{\prime}$, and $H_8^{\prime}$ is a parameter that describes the
$g \rightarrow \chi$ fragmentation $via$ a $S$-wave $c\bar c$
colour-octet state.  We assume that the $H_1$ for the $2P$ states is
about the same magnitude as that of the $1P$ states
\cite{isgur1}.  The parameter $H_8^{\prime}$ has a large range of
uncertainty even for the $1P$ states.
On general grounds one expects the $H'_8$ parameter for the $2P$ state
to lie in the same range \cite{close}.  Firstly the $H_1$ and $H'_8$
contributions to any $P$-wave quarkonium production are expected to be
comparable in size.  Secondly the wave function factor $R_0^2$, which
is the colour singlet analogue of $H'_8$, is similar in size for the
$1S$ and $2S$ states.  Therefore we assume the same $H'_8$ parameter
for the $2P$ state as for $1P$;
whether or not they are exactly equal is immaterial in view of the
large uncertainty in this parameter.  The parameter values used are
$R_0^2 = 0.49~{\rm GeV}^3$ \cite{schuler}, $H_1 = 15.0~{\rm MeV}$
and $H'_8= 3.0~{\rm MeV}$ \cite{bryu,bryu2}.
 We take the masses of the $2^3P_1$ and the
 $2^3P_2$ states to be 3.95~GeV and 3.98~GeV, respectively
\cite{isgur1}. In our computations, we use \cite{plothow} the updated
MRSD-${}^{\prime}$ parametrisations \cite{mrs} for the parton
densities in the nucleon, evolved to a scale $Q^2=\mu^2/4$, where
$\mu$ is chosen to be $M_T$ for the fusion process, and equal to
$p_T^{g,c}=p_T/z$ for the fragmentation process.  The fragmentation
functions are evolved to the scale $p_T/z$.

\vskip10pt
Finally, we come to the branching ratios for the radiative decay
$2^3P_{1,2} \rightarrow 2^3 S_1 + \gamma$.  The absolute value of
these widths have been estimated by rescaling the known values for the
bottomonium system \cite{close}.  One gets $\Gamma(2^3P_{1,2}
\rightarrow 2^3 S_1 + \gamma) \sim 0.1 - 0.2~{\rm MeV}$.  The
corresponding total widths have been estimated, assuming phase-space
and dynamical suppressions into open charm states, to be $\sim 2 - 5$
MeV each \cite{close}.  The resulting branching ratios for $2^3P_{1,2}
\rightarrow 2^3 S_1 + \gamma$ lie in the range 2 -- 10\%.  The results
presented below are based on the optimistic value of the branching
ratios = 10\%.

\vskip10pt
In Fig.~1 we present our results for $Bd\sigma /d p_T$ for
$\psi^{\prime}$ production in $\bar p p$ collisions at $\sqrt{s}=
1.8$~TeV and integrated over a pseudo-rapidity interval $\vert
\eta \vert \le 0.5$, and compare our predictions to the data from the
CDF experiment \cite{cdf}. $B=7.7 \times 10^{-3}$ is the $\psi^{\prime}$
branching ratio into leptons. As in the case of $J/\psi$ production
\cite{ours}, we find the $g \rightarrow P$-state fragmentation
contribution to be dominant. The inclusion of the contribution
of the $2P$ states increases the $\psi^{\prime}$ cross-section
significantly, and assuming a value of 10\% for the branching ratios
this is capable of explaining the bulk of the observed cross-section.
It should be noted of course that this choice of the $2P \rightarrow
2S$ branching ratios corresponds to the upper end of the estimated
range.  Within the lattitude of this range the cross-section can go
down by a factor of 5.  Further there is an uncertainty of a factor of
3 on either side in the production cross-section of the $2P$ states,
arising largely from the $H_1$ and $H'_8$ parameters.  Consequently
the predicted normalisation of the $\psi'$ cross-section can be
anywhere between the experimentally observed range and an order of
magnitude below. Thus it is premature to rule out the necessity of
more exotic contributions, e.g. from the hybrid states suggested in
\cite{close}.  It is remarkable nonetheless that the inclusion of
the $2P$
contributions can push up the predicted $\psi'$ cross-section by the
required factor of $\sim 30$ within the lattitude of the model parameters.

\vskip10pt
In conclusion, we find the $2^3P_{1,2}$ contribution to
$\psi^{\prime}$ production at CDF to be very important. We have computed
the fusion and fragmentation contributions to these $2P$-states and
to the $2S$ $\psi^{\prime}$ state, and find that with an optimistic
choice of the $2P \rightarrow 2S$ branching, it is
possible to explain the anomalously large cross-section for
$\psi^{\prime}$ production reported by at the CDF experiment.

\vskip10pt
We are grateful to Frank Close for communicating his work of Ref.
\cite{close} to us before its public distribution.

\newpage
\section*{Figure captions}
\renewcommand{\labelenumi}{Fig. \arabic{enumi}}
\begin{enumerate}
\item
The cross-section $Bd\sigma/dp_T$ (integrated over the pseudorapidity
range $-0.5< \eta < 0.5$) for the process $\bar p p \rightarrow
\psi^{\prime} X$ as a function of $p_T$ at $\sqrt{s}=1.8$~TeV. The
data are taken from Ref.~\cite{cdf}.  The different curves correspond
to the direct production via fusion (dashed line), the gluon
fragmentation contribution (dashed-dotted line), the charm quark
fragmentation term (dotted line) and the sum of all contributions
(solid line).
\end{enumerate}
\end{document}